\documentclass[final,5p,times,twocolumn]{elsarticle}

\usepackage{amssymb}
\usepackage{amsthm,amsmath}

\journal{Physics Letters B}

\begin{document}

\begin{frontmatter}



\title{Disk galaxies and their dark halos as self-organized patterns}

\author[label1]{Shankar C. Venkataramani\corref{cor1}}
\cortext[cor1]{Corresponding author}
\ead{shankar@math.arizona.edu}
\author[label1]{Alan C. Newell}
\ead{anewell@math.arizona.edu}
\address[label1]{Department of Mathematics, University of Arizona, 617 N. Santa Rita Ave., Tucson, AZ 85721, USA}



\date{\today}

\begin{abstract}
Galaxies  are built by complex physical processes with significant inherent stochasticity. It is therefore surprising that the inferred dark matter distributions in galaxies are strongly correlated with the observed baryon distributions leading to various `Baryon-Halo conspiracies'. The fact that no dark matter candidate has been definitively identified invites a search for alternative explanations for such correlations and we present an approach motivated by the behaviors of self organized patterns. We propose a nonlocal relativistic Lagrangian theory for a `pattern field' which acts as an `effective dark matter', built on the idea that defects in this pattern field couple to the baryonic matter distribution.
Our theory accounts for the gross structure of cold disk galaxies. We compute galactic rotation curves and derive various galaxy scaling relations including Renzo's rule, the radial acceleration relation, and  the existence of the Freeman limit for central surface brightness. 
\end{abstract}




\end{frontmatter}





%
%

\section{Mass discrepancies and dark matter}

Understanding what we call dark energy and dark matter are two of the great intellectual challenges of our time. Each is a place-holder for current ignorance. Dark energy is postulated to explain why the expansion of our universe is  accelerating, and dark matter is postulated to explain discrepancies between the observed (non-relativistic) motions of stars in galaxies, and galaxies in clusters, from what one would predict using Newtonian gravity \cite{Trimble1987existence}. Although long suspected (cf. \cite{Oort1932force,Zwicky1933Rotversciebung}), the  hunt for additional `invisible' matter  became a serious endeavor in the wake of the pioneering observations by Vera Rubin and her colleagues \cite{Rubin_Rotation_1970,Rubin_Extended_1978},
 demonstrating definitively that the rotation velocity curves $v(r)$ of galaxies flatten out with increasing radius $r$ instead of the expected Keplerian decay $v \approx \sqrt{GM/r}$ predicted by a balance of the gravitational $GM/r^2$ and centrifugal $v^2/r$ accelerations. 

The dominant paradigm in cosmology is that the universe began with a big bang, it has nearly critical density $\Omega \approx 1$, dark energy 
is modeled by a non-zero cosmological constant $\Lambda$, 
and the bulk of the matter in the universe consists of (dynamically) cold dark matter (CDM) that  clumped into halos and seeded the formation of galaxies and larger scale structures in the universe \cite{Ostriker1995Observational}. There are multiple lines of evidence supporting this theory including the presence of the cosmic microwave background (CMB), the relative (primordial) abundances of the light elements (cf. \cite{Peebles1966Primeval}), the spectrum of temperature anisotropies in the CMB \cite{Hu1997physics}, 
the large-scale distribution of matter in the universe, 
and the observed acceleration in the expansion of the universe \cite{Perlmutter1999measurements,Riess1998Observational}.

Our focus in this letter is on the structure and dynamics of disk galaxies. In the conventional picture, disk galaxies have two distinct components -- a massive `CDM halo', which is roughly spherical, and a thin disk, containing stars/gas (baryons) that is in rotational equilibrium in the combined gravitational field of the halo and the disk. 

We now discuss each of these components in turn.
While the CDM model for dark matter works remarkably well on 
cosmological scales, no DM particle has yet been definitively identified. Additionally, there are several discrepancies between CDM predictions and observations on galactic or smaller scales \cite{deMartino2020Dark}. $N$-body simulations give spherical halos with the ``universal" Navarro-Frenk-White (NFW) profile $\rho_{\text{NFW}}(R) = \frac{\rho_0 }{ [(1+(R/R_S)^2)(R/R_s)]}$ \cite{NFW96}, ($R$ is the 3d radial coordinate). Observations, however, favor ``cored" halos, e.g. the quasi-isothermal profile $\rho_{\text{qiso}} = \frac{\rho_0}{(1+R^2/R_C^2)}$, over the ``cuspy" NFW profile \cite{Li2020Comprehensive}. 

 A proxy for the distribution of baryonic matter in disk galaxies is given by the surface brightness profile. In the disks of galaxies, i.e. outside the bulge if one is present, the brightness decays (approximately) exponentially from the center \cite{Freeman1970disks}. Assuming a constant mass-to-light ratio, the baryonic surface density $\Sigma = \Sigma_0 \exp(-r/r_0)$, where $r$ is the 2d radial coordinate in the galactic plane, $\Sigma_0$ is the central surface density and $r_0$ is the (baryonic) scale length of the galaxy. The Freeman ``law" is observational evidence that $\Sigma_0 \approx \Sigma^*$ is the same for all high surface brightness (HSB) galaxies, {\em independent of their total mass}   \cite{Freeman1970disks}. Including low surface brightness (LSB) galaxies gives a wider distribution of central densities, with a rapid fall-off beyond $\Sigma^*$, defining the {\em Freeman limit} \cite{McGaugh1995galaxy}. 
 
 Observations reveal  tight correlations and scaling relations between the halo parameters $\rho_0,R_C$ and the baryonic parameters $\Sigma_0, r_0$ \cite{Donato2004Cores,Donato2009constant}. Indeed, galaxies are {\em surprisingly simple} and 
 seemingly governed by a single dimensionless 
 parameter  \cite{Disney2008Galaxies}. Since galaxy formation is inherently stochastic this suggests an important role for self-organizing dynamical processes \cite{Aschwanden2018order}. 
 
 For quasi-steady systems, many observations indicate that the dynamically inferred DM halo is strongly correlated with the baryon distribution \cite{Famaey2012MOND}. Many of these relations are subsumed by the radial acceleration relation (RAR) \cite{Lelli2017onelaw}, which is a ``local" relation for the observed total acceleration  $g_{\text{obs}}$ (from halo + disk) and the purely baryonic contribution $g_{\text{bar}}$.  This relation holds for a range of galaxies including dSphs, disk galaxies (S0 to dIrr) and giant ellipticals, and was first proposed in \cite{Milgrom_MOND_1983} as the basis for `Modified Newtonian dynamics (MOND)'. The successes of MOND in predicting various observed regularities 
on galactic scales \cite{McGaugh2020predictions} have inspired a variety of dark matter models that behave like MOND for galaxies and like $\Lambda$CDM on cluster and larger scales \cite{Zhao2010Dark,Khoury2015Alternative}. There have also been attempts to recover these scaling relations within $\Lambda$CDM using  cosmological simulations that include various baryonic feedback mechanisms \cite{Navarro2017MDARfull,Dutton2019origin}. 

Our aim in this letter is to propose specific self-organizing mechanisms for galactic dynamics. To this end, we present a novel theory that allows us to compute rotation curves of galaxies, and explain many of the observed galaxy scaling relations. We derive the RAR and the Freeman limit from the energy of defects in a pattern associated with the self-organizing mechanisms in galaxies. We are motivated in this endeavor by the self-organizing properties of pattern forming systems and a recognition that instability generated patterns have a role to play in galaxies  \cite{Aschwanden2018order}. In many situations, pattern structures are topologically constrained by the presence of defects and are not ground states. Therefore they can store energy.  Let us emphasize this paradigm changer. {\em Additional energy that depends on the baryonic matter distribution can give rise to forces that produce effects attributed to dark matter halos}. In such a scenario, it is not at all surprising that there should be tight correlations and scaling relations between halo and baryonic parameters.

%
%

\section{Patterns, universality, defects and halos}

Patterns are ubiquitous in nature and arise when, at some stress threshold, a symmetric ``ground state" destabilizes and certain symmetry-breaking modes are preferentially amplified. These modes compete for dominance through nonlinear interactions and a set of winning 
configurations emerges. Generally, whereas some symmetries are broken, others are not, leading to the presence of defects that prevent the new state from being a ground state, a true energy minimum.

A useful illustration is provided by the well-studied case of high Prandtl number convection in a horizontal layer of fluid heated from below. For a sufficiently large thermal gradient, the conduction state becomes unstable to convective rolls which transport heat more efficiently. At this transition the continuous translation symmetry of the conductive state is broken and replaced by a discrete translation symmetry from the preferred wavelength of the roll pattern. However, because the rotational symmetry is not broken at the transition, the orientations of the roll patches are chosen by local biases, boundary conditions and other constraints. If the system size is much greater than the chosen wavelength, the resulting pattern is a mosaic of ``locally" uniform stripe patterns with different orientations which meet and meld along defect lines and points in 2D (and planes and loops in 3D). In more confined geometries such as cylinders or spheroids where the boundaries may be heated, or in situations where angular momentum conservation constraints might apply, the patterns, although locally stripe-like, can be target or spiral shaped. The resulting defects have topological charges reflecting the far-field geometry or constraints away from the defects. They also have energy, associated with the fact that the emerging pattern is not a
true energy minimum but a metastable state; metastable in the sense that either the topological constraints make the state a local minimum or that the time scale to coarsen and ``heal" the
defects is extremely long.

Patterns and other collective phenomena are studied using macroscopic {\em order parameters} that measure the amount of symmetry breaking. Order parameters are governed by universal equations that reflect the underlying symmetries of the system but are insensitive to the precise details of the microscopic interactions in the system -- a phenomenon called {\em universality} \cite{Kadanoff1990scaling}. For systems that form stripe patterns by breaking translation but not orientation invariance, the appropriate order parameter is a phase $\psi$ whose gradient $\mathbf{k}$ gives the local orientation of the pattern. The microscopic fields are generally $2\pi$ periodic functions of the phase $\psi$. Integrating over the microscopic degrees of freedom gives a canonical form for the  effective energy \cite{NV17}
\begin{equation}
    \label{Energy_RCN} \mathcal{E} = \rho_0 c^2\int \left[(1 - \mathbf{k}^2)^2 + k_0^{-2}(\nabla \cdot \mathbf{k})^2  \right] dV, \quad \mathbf{k} = \nabla \psi,
\end{equation}
where $\psi$ is a length, the wavevector $\mathbf{k}$ is dimensionless, and the normalizing constant $\rho_0 c^2$ ensures dimensional consistency. The ground states $\mathcal{E} = 0$ correspond to the plane waves, $\psi(\mathbf{x}) = \mathbf{k}\cdot \mathbf{x}$ at spatial location $\mathbf{x}$, with the preferred wavelength $|\mathbf{k}| = 1$. If boundaries or other external constraints dictate that the phase pattern be radial, $\psi(\mathbf{x}) = \psi(R)$ where $R = |\mathbf{x}|$, we cannot be in a ground state. Indeed, a calculation reveals that, minimizing $\mathcal{E}$ with $\psi = \psi(R)$,  we get $\mathbf{k} \to 0$ as $R \to 0$ and $\psi(R) \to R + \text{const}$ as $R \to \infty$. 
These target (spirals if the instability is to waves and~\eqref{Energy_RCN} is modified appropriately) patterns are robust because they cannot be continuously deformed into the plane wave ground states.
Their curvature radii are large compared to the local pattern wavelength, so they are locally stripe like and their macroscopic energies can be represented by~\eqref{Energy_RCN}.

What is important to us here is that the target pattern has an energy density of the same form as  a cored quasi-isothermal halo with $R_C \sim k_0^{-1}$. Such halos describe the dark matter distribution out to the edge of the optical disk in real galaxies,  so this suggests adding a term like~\eqref{Energy_RCN} to the Lagrangian of a galaxy can recover the effects of `dark matter' \cite{NV17,Newell2019pattern}. And indeed, it does! The additional action leads to the prediction of many of the features observed in galaxies, 
including the flattening of rotation curves 
to a limiting value 
$v_\infty^2  =\frac{\sqrt{32\pi G \rho_0}}{k_0}$ \cite{Newell2019pattern}.
While these results were encouraging important questions were left unanswered - (1) What determines the parameters $\rho_0$ and $k_0$?,  (2) How do we eliminate the assumption of spherical symmetry and model more realistic disk galaxies? and, (3) What physical processes might lead to a term like~\eqref{Energy_RCN} in the action?
We address the first two questions in subsequent sections and demonstrate therein that our framework captures the observed phenomenology of disk galaxies very accurately. We suggest candidates for the last challenge (3) and discuss potential future directions in the concluding remarks.

%
%

\section{An effective Lagrangian for pattern dark matter}

A proper formulation of our ideas requires the identification of an appropriate Lagrangian. The Eintein-Hilbert Lagrangian for the geometry of spacetime, along with the `dust' Lagrangian for CDM/baryons, and the cosmological constant for dark energy, describes $\Lambda$CDM. Alternative Lagrangians describe various flavors of MOND \cite{Beckenstein1984does,Beckenstein2004TeVeS,Skordis2020RelMOND}, dark matter with novel material properties -- dipolar dark matter \cite{Blanchet2017Dipolar}, superfluid dark matter \cite{Berezhiani2015theory} and fuzzy dark matter \cite{Hu2000fuzzy}, as well as ``non-material" alternatives like emergent gravity \cite{Hossenfelder2017Covariant}. Our goal is to formulate an appropriate `pattern dark matter' Lagrangian that encapsulates the physics discussed above. 

The key  ideas underlying our approach are empirically derived from observations. They are: 
\begin{enumerate}
\item The stellar and halo scale-lengths of disk galaxies, $r_0$ and $R_C$,  are characterized by a single length scale \cite{Donato2004Cores}, 
\item Disk galaxies have a universal acceleration scale $a_0  \approx 1.2 \times 10^{-10} \mathrm{m/s}^2$ \cite{Milgrom_MOND_1983}. Fitting the rotation curves using quasi-isothermal halos gives a universal central surface density $\rho_0/k_0 \simeq \Sigma^* \sim 100 \,M_{\odot}/\text{pc}^2 \approx a_0/(2 \pi G)$ \cite{Donato2009constant,Famaey2012MOND}, 
\item The Baryonic Tully-Fisher relation $v_\infty^4 = GM_Ba_0$ (BTFR) relates the baryonic mass $M_B$ to $v_\infty$ with very little scatter over a wide range of galaxies \cite{McGaugh2000BTFR},  and
\item The dark matter halo couples to the local baryonic density $\rho_B$, and not just the total mass $M_B$ \cite{Sancisi2004visible}. 
\end{enumerate}

The constancy of core surface density of halos, and the BTFR,   
(points (2) and (3) in our list) give
\begin{align}
\label{calibration}
 \frac{\rho_0}{ k_0} =  \Sigma^*, \quad v_\infty^2 = \frac{32 G \pi \rho_0}{k_0^{2}} & = \sqrt{G M_B \Sigma^*}. \end{align}

Based on these observations, we now propose a Lagrangian theory for the coupled baryon-pattern system where the energy stored in the pattern is effectively {\em ``Dark Matter"}.  
Our action is given by $\mathcal{S} = \mathcal{S}_{EH}  + \mathcal{S}_M  + \mathcal{S}_P  + \mathcal{S}_{\psi}$, with
\begin{align}
\mathcal{S}_{EH} & = \frac{c^4}{16 \pi G}\int R \sqrt{-g} \, d^4 x, 
\quad \mathcal{S}_M   = \int \rho_B u^\alpha u_\alpha \sqrt{-g} \, d^4 x, \nonumber \\
\mathcal{S}_P&  = -\rho_0 c^2  \int \left\{(1-\nabla^\mu \psi \nabla_\mu \psi)^2 +  k_0^{-2}(\nabla^\mu \nabla_\mu \psi )^2\right\} \sqrt{-g} \ d^4x, \nonumber \\
{\mathcal S}_{\psi} & = -\int \rho_B c^2 V[\nabla^\beta \psi \nabla_\beta \psi] \sqrt{-g} d^4 x, \nonumber \\
\rho_0 & = 32 \pi {\Sigma^*}^{3/2} (G M_B)^{-1/2}, \quad k_0 = 32 \pi {\Sigma^*}^{1/2}{(G M_B)}^{-1/2},\label{Lagrangian}
\end{align}
where $\rho_0,k_0$ are obtained by solving~\eqref{calibration}, $\nabla$ represents the covariant derivative in a curved spacetime and $u^\alpha$ is the 4-velocity of the baryonic matter. 

All the terms in the Lagrangian are well motivated. The first two pieces are canonical choices --  the Einstein-Hilbert action $\mathcal{S}_{EH}$ and the matter action $\mathcal{S}_M$ for dust. 
The third term $\mathcal{S}_P$ is given by the pattern Lagrangian for curved spacetimes, as in~\eqref{Energy_RCN}. The last term $\mathcal{S}_\psi$ models a direct interaction between baryonic density $\rho_B$ and the pattern field $\psi$. $V(|\mathbf{k}|^2) \geq 0$ is a convex potential that vanishes at $\mathbf{k} = 0$, so large $\rho_B$ creates ``defects" in $\psi$, like the spherical target pattern with $\nabla \psi = 0$ at the center.
 $\mathcal{S}_P$ and $\mathcal{S}_\psi$ come with negative signs since they are `potential'  terms, i.e. akin to $V$  in the action $S = \int (T-V) dt$. 

Since galaxies are non-relativistic, $v_{\infty} \ll c$, 
the geometry of space-time deviates from the flat Minkowskii space at $O(\epsilon)$ where 
$\epsilon=\left(\frac{v_\infty}{c}\right)^2.$ 
We obtain the (Newtonian) limit description through a principled asymptotic expansion in the small parameter $\epsilon$. In a steady state, our system is described by the weak-field metric, 
$
g  = -(c^2+ 2 \phi(\mathbf{x})) dt^2 + (1 - 2\phi(\mathbf{x})/c^2) (dx^2+dy^2+dz^2), 
$ where $\phi(\mathbf{x})$ is the total Newtonian potential. We note that $\psi, \mathbf{x}, \mathbf{k}$
are $O(1)$, the spatial velocity $\mathbf{v} = \frac{d\mathbf{x}}{dt}$ is $O(\sqrt{\epsilon})$, and $\rho_B$, $\rho_0$ and $\phi$ are
$O(\epsilon)$.
We can expand the action $\mathcal{S}$ and collect terms in powers of $c$ (equivalently $\epsilon$) to get,  $\mathcal{S} = c^2 \mathcal{S}_1 + \mathcal{S}_2$,
\begin{align}
       \mathcal{S}_1 & = -\int d^3\mathbf{x} dt\,\left[ \rho_0 [(1-|\nabla \psi|^2)^2 + k_0^{-2}(\Delta \psi)^2] + \rho_B V(|\nabla \psi|^2) \right] \nonumber \\
    \mathcal{S}_2 & = \int  d^3\mathbf{x} dt \, \left[\rho_B \left(\frac{\mathbf{v}^2}{2} -\phi\right) -\frac{|\nabla \phi|^2}{8 \pi G} - 2 \phi \rho_0  \left(|\nabla \psi|^4-1 \right)  \right. \nonumber \\
    & - \left. 2 \phi\left(\rho_0 k_0^{-2}  \left(\Delta \psi\right)^2 + \rho_B V'(|\nabla \psi|^2) |\nabla \psi|^2 \right)\right].
    \label{newtonian-limit}
\end{align}
This formulation is completed by prescribing the potential $V$.

%
%

\section{Phase surfaces as dark halos}

We illustrate the procedure for analyzing the variational equations for the action in~\eqref{Lagrangian} by revisiting the example of spherically symmetric compact clump of matter. {\bf Step 1:} Prescribe $\rho_B(R)$ and solve the variational equations for $\mathcal{S}_1$, i.e. a pattern formation problem. For a compact clump, and a generic potential $V$ with a global minimum at 0,  $\nabla \psi \approx 0$ within the source, so we get the target patterns that were discussed earlier. {\bf Step 2:} With the given $\rho_B$  and $\psi$ computed from the previous step, solve for the gravitational potential $\phi$. For a compact dense clump, $\nabla \psi\approx 0$ where $\rho_B \neq 0$, and outside the clump, $|\nabla \psi| \approx 1, \Delta \psi \approx 2 R^{-1}$. Consequently, we get $\Delta \phi \approx 4 \pi G(\rho_B + 8 \rho_0/ (b^2 + k_0^2 R^2)), b\sim O(1)$, 
\begin{equation}
g = \nabla \phi \approx \frac{GM_B}{R^2} + \frac{32 \pi G \rho_0}{k_0^2 R} \left[1- \frac{b}{k_0 R} \tan^{-1}\left(\frac{k_0R}{b}\right)\right].
\end{equation}
{\bf Step 3:} Solve for the steady state velocity from $\frac{v^2}{R} = g$. 

\begin{figure}
  \centering
  \includegraphics[width=0.5 \textwidth]{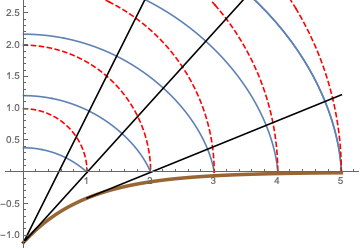} 
  \caption{Huygens' construction -- 
  the phase contours (solid curves in $z > 0$) have a common evolute (solid curve in $z<0$)  and intersect the characteristics (straight lines) orthogonally. 
  The  contours for $z \leq 0$ are given by reflection. 
  A spherical target pattern (dashed phase contours) is shown for comparison. The involutes are (approximately) spherical caps with centers off the plane $z=0$.
      }
  \label{construction}
\end{figure}

To model a disk galaxy, we now carry out these steps in an axisymmetric setting, where all the fields only depend on $r = \sqrt{x^2+y^2}$ and $z$. The galactic disk is on the plane $z=0$ and the matter density is concentrated close to this plane, so we take $\rho_B(r,z) = \Sigma_B(r) \delta(z)$. 
 
 In Step 1, extremizing $\mathcal{S}_1$, we have two contributions, the pattern Lagrangian $\mathcal{S}_P$ which is an integral over all of space, and the interaction Lagrangian $\mathcal{S}_\psi = -2 \pi \int \Sigma_B(r) V(|\psi_r|^2) r dr$ which is an integral over the galactic disk. Off the disk  $\psi$ satisfies the {\em Eikonal equation} $|\nabla \psi| = 1$, as appropriate for stripe patterns. 
Using  Huygens' principle, we obtain:
 \begin{align}
& \psi(r,z) = \min_{s \geq 0} \left[\psi(s,0) + \sqrt{(r-s)^2 + z^2}\right]  \nonumber \\
 \Rightarrow \quad & \psi\left[s+  t \cos \theta(s), \pm t \sin \theta(s) \right]  = \psi(s,0) + t. 
 \label{huygens}
\end{align}
where the second line follows for regions where the  {\em characteristics} $r = s+  t \cos \theta(s), z=\pm t \sin \theta(s))$ do not cross. 

\begin{figure}
  \centering
  \includegraphics[width=0.45 \textwidth]{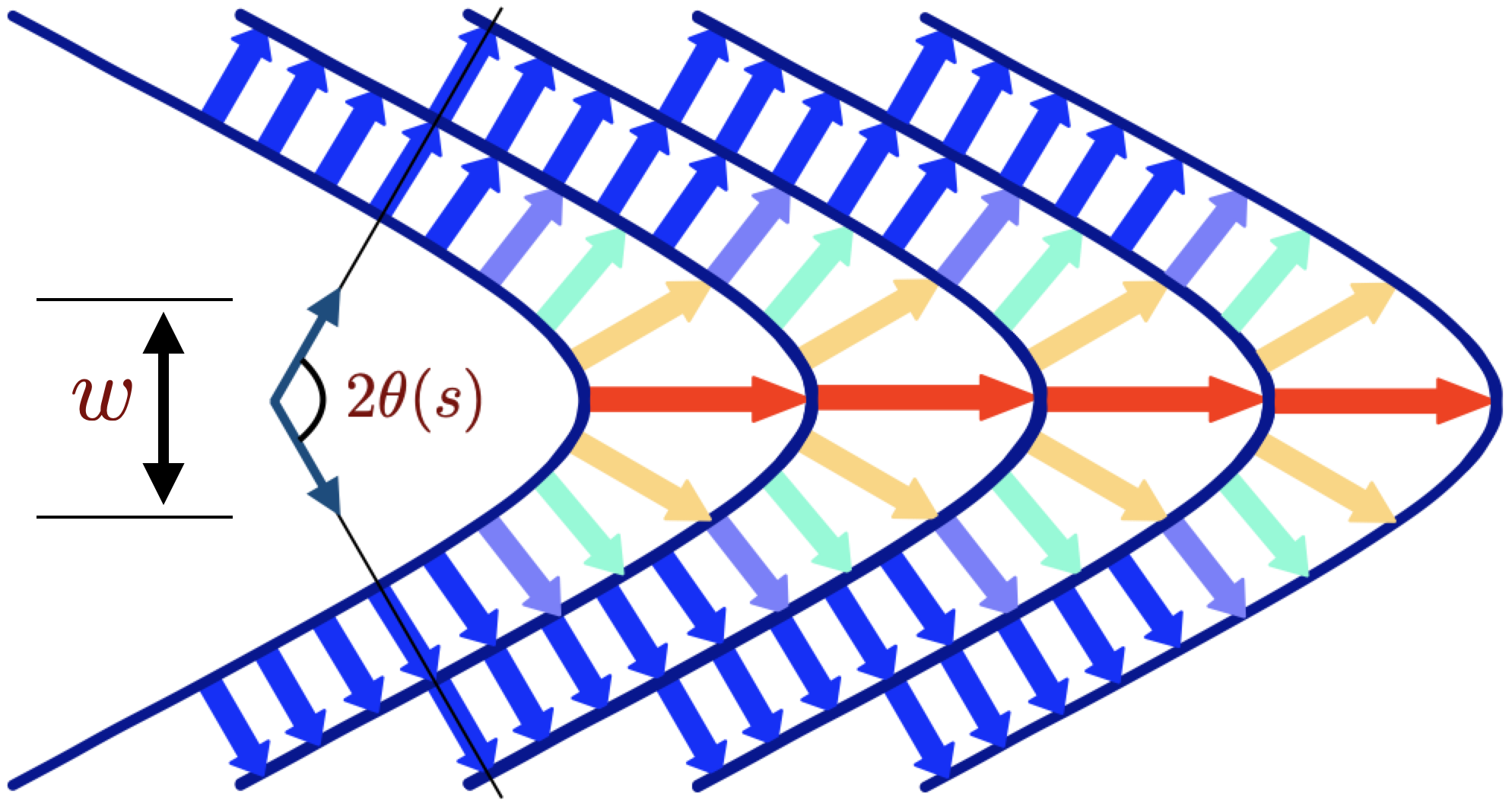} 
  \caption{Phase grain boundary (PGB). There is a jump in $\nabla \psi$ across the PGB. This structure is smooth on the scale $w$, the width of the PGB. The stretching and bending of the phase contours  contribute to an effective surface energy.}
  \label{fig:PGB}
\end{figure}

The geometry of this construction is illustrated in Fig.~\ref{construction}. The phase fronts  for $z > 0$ (resp. $z < 0$) are the {\em involutes} of a common {\em evolute} $\gamma = (\alpha(s),\mp \beta(s))$ and the phase $\psi$ is the local radius of curvature \cite[\S 12]{Rutter2000Geometry}. For the Eikonal solution, $\nabla \psi$ is discontinuous across the galactic plane $z=0$. Indeed, in contrast to the spherical target pattern, the contours given by the involutes intersect the plane $z=0$ at an angle $\theta(s) \neq \frac{\pi}{2}$. This discontinuity in $\nabla \psi$ is regularized as a {\em phase grain boundary} (PGB), a defect well known in patterns, consisting of  a boundary layer across which  $\nabla \psi$ changes smoothly as illustrated in Fig.~\ref{fig:PGB}.  

We can estimate the (surface) energy density of a PGB as follows. Since the boundary layer has width $w$, the curvature and stretch of the phase contours are, respectively, 
$\Delta \psi \sim w^{-1} \sin \theta(s), 1-|\nabla \psi|^2 \sim \sin^2 \theta(s)$.
Eq.~\eqref{Energy_RCN} now implies  
$$
\Sigma_{PGB} \sim w \left\{(\Delta \psi)^2 + (1-|\nabla \psi|^2)^2)\right\}  \sim w^{-1} \sin^2 \theta(s) + w \sin^4 \theta(s).
$$ 
Optimizing for $w$ gives $w \sim \frac{1}{\sin \theta(s)}, \Sigma_{PGB} \propto \sin^3 \theta(s)$. A rigorous calculation along these lines yields $\Sigma_{PGB} = \frac{8 \rho_0}{3k_0} \sin^3\theta(s)$ \cite{newell1996defects}. Using~\eqref{calibration}, the sum of $\mathcal{S}_\psi$ and the PGB defect energy is 
\begin{equation}
    \label{disk-energy}
    \mathcal{S}_{\text{disk}} = 2 \pi \int \left[\frac{8 \Sigma^*}{3} \sin^3 \theta(s) + \Sigma_B(s) V(\cos^2\theta(s)) \right] s ds.
\end{equation}
We can extremize to get $\Sigma_B(s) V'(\cos^2 \theta(s))=4 \Sigma^* \sin \theta(s)  $, {\em a local relation} between the matter surface density, the characteristic angle $\theta(s)$, and indirectly, also the common evolute $\gamma$.

We can now make an informed choice for the potential $V$. The argument of $V$ is $| \nabla \psi|^2 =\cos^2 \theta(s) \leq 1$ within the galactic disk. To ensure $|\nabla \psi|^2 \leq 1$ in the presence of matter, a canonical choice is the {\em log barrier function} $V = -V_0 \ln(1-|\nabla \psi|^2)$ \cite{boyd2004convex}, where $V_0$ is an $O(1)$ constant. Putting everything together, we have the leading order (in $\epsilon = (v_\infty/c)^2$) solution of the variational equations for~\eqref{Lagrangian}:
\begin{align}
(r,z) & = (s + t \cos \theta(s), \pm t \sin \theta(s)), \nonumber \\
\gamma & = \left(s + \frac{\cos \theta(s) \sin \theta(s)}{\theta'(s)}, \frac{\sin^2 \theta(s)}{\theta'(s)}\right) \nonumber \\
    \Sigma_B(s) & = \frac{4 \Sigma^*}{V_0} \sin^3 \theta(s) \nonumber \\
    |\nabla \psi| & \simeq 1, \quad \Delta \psi \simeq 2 \left(t - \sin \theta(s) /\theta'(s)\right)^{-1} \approx 2/\sqrt{r^2+z^2}, \nonumber \\
    \Delta \phi & = \Delta (\phi_{B} + \phi_P) \simeq  4 \pi G \left[\Sigma_B(r) \delta(z) + 2 \rho_0 k_0^{-2} (\Delta \psi)^2\right], \nonumber \\
    v^2 & =  r  \partial_r \phi(r,0) =r  \partial_r \phi_B(r,0) + r \partial_r \phi_P(r,0).
    \label{galaxy}
\end{align}

We record a few observations. Our equations describe equilibria 
for {\em purely rotation supported}  galaxies (no significant bulge or pseudobulge). Such galaxies satisfy $\Sigma_B \leq \frac{4 \Sigma^*}{V_0}$, so the Freeman limit \cite{McGaugh1995galaxy} 
follows naturally from our analysis. 

The second equation expresses the common evolute $\gamma$ in terms of $\theta(s)$ which in turn is given by $\Sigma_B$. This gives a direct link between the {\em local} matter distribution $\Sigma_B$ and the pattern `halo', a quantitative formulation of Renzo's rule -- features in the light distribution are reflected in the rotation curves \cite{Sancisi2004visible}. 

We can also prescribe $\gamma$ and use it to compute $\Sigma_B, \psi,\phi$ and $v$. A natural {\em critical} case is when the evolute degenerates to a single point $(0,-z_0)$, so that $\theta(s) = \arctan(\frac{z_0}{s})$ and $\Sigma_B(s) = \frac{4\Sigma^*}{V_0}(1 + s^2/z_0^2)^{-3/2}$, corresponding to a Kuzmin disk. 

It is remarkable that the surface density of a Kuzmin disk follows directly from surface energy $\propto \sin^3\theta(s)$ relation for PGB defects, a formula that was originally derived in the context of patterns  \cite{newell1996defects}. The mass of this `critical' Kuzmin disk, $M_B = 8 \pi \Sigma^* z_0^2/V_0$, is determined by $z_0$, the length-scale in the evolute. The phase is given by $\psi(r,z) = (r^2 + (|z| + z_0)^2)^{1/2}$. We can compute the potential $\phi(r,0)$ 
and the rotation velocity:
\begin{align}
    \label{kuzmin-pot}
    \frac{\phi(r,0)}{2 \pi G \Sigma^* z_0} & \approx -\frac{4}{V_0\sqrt{1+\xi^2}}  +  V_0^{-1/2} \log(1+\xi^2) + \cdots, \nonumber \\ 
    \frac{v^2(r)}{2 \pi G \Sigma^* z_0}  & \approx \frac{4 \xi^2}{V_0(1+\xi^2)^{3/2}} + 2 V_0^{-1/2}\frac{\xi^2}{{1+\xi^2}} + \cdots,  \nonumber \\
    \frac{g_{\text{obs}}}{2 \pi G \Sigma^*} & \approx \frac{4 \xi}{V_0(1+\xi^2)^{3/2}} + 2 V_0^{-1/2}\frac{\xi}{{1+\xi^2}} + \cdots 
\end{align}
where $\xi = r/z_0$ is the scaled radius, and the initial terms are the (non-dimensional) baryonic contributions to the potential ($\phi_{\text{bar}}$), velocity ($v_{\text{disk}}$) and acceleration ($g_{\text{bar}}$).  The asymptotic velocity $v_\infty^2 = 4 \pi V_0^{-1/2} G \Sigma^* z_0 \equiv (G M_B a_0)^{1/2}$ where $a_0 = 2 \pi G \Sigma^*$. We obtained $\rho_0$ and $k_0$ for a spherical halo, so it is promising that our theory recovers the BTFR in a different regime with a spheroidal halo and an extended, disk like matter distribution. Also, independent of the scale $z_0$, the critical Kuzmin disks in our theory satisfy a radial acceleration relation (RAR) since both $\frac{g_{\text{bar}}}{a_0}$ and $\frac{g_{\text{obs}}}{a_0}$ only depend on the combination $\xi = r/z_0$. 

\begin{figure} 
  \centering
  \includegraphics[width=0.5 \textwidth]{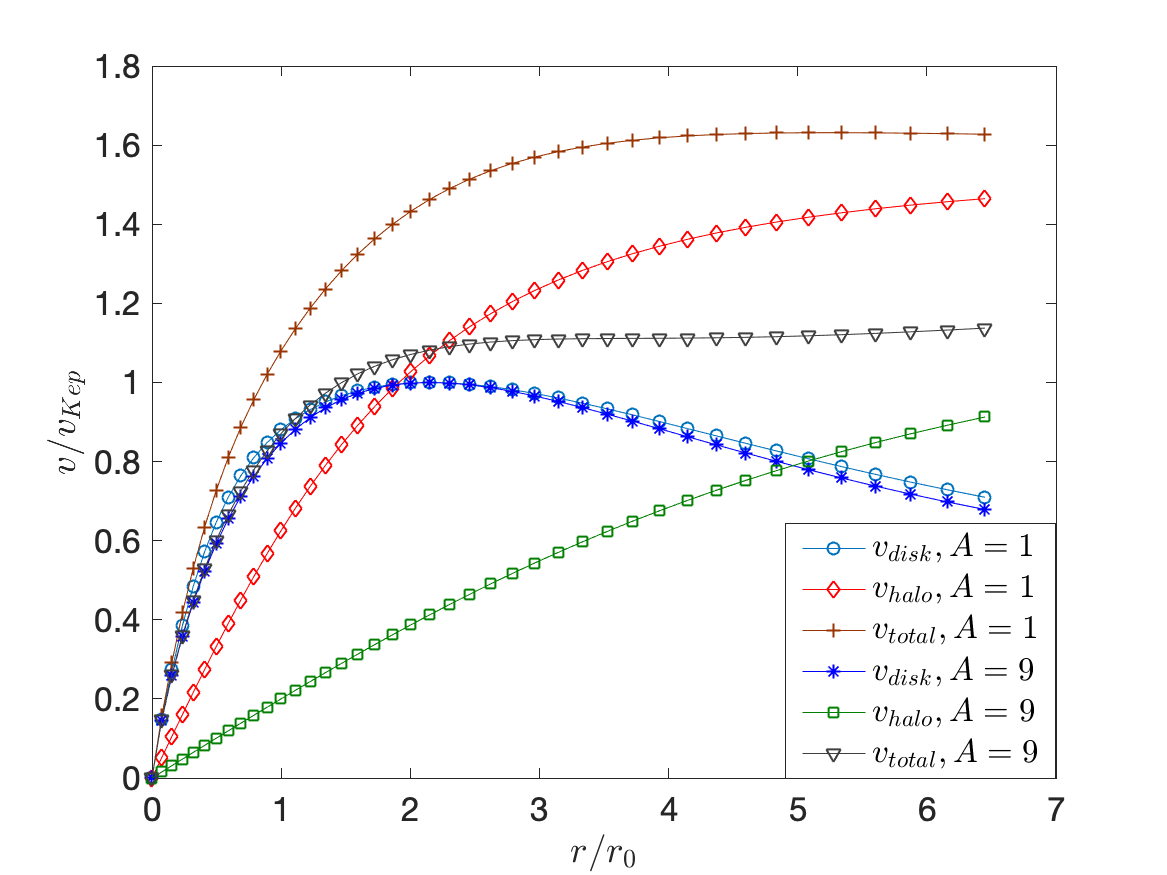} 
  \caption{   Computed rotation curves. We approximate the 
    pattern DM $2 \rho_0 k_0^{-2} (\Delta \psi)^2$ by the $\ell = 0$ mode and compute the disk potential from a minimum discrepancy decomposition into Exponential + Kuzmin disks.  Compare Fig.~7 in Ref.~\protect{\cite{DiPaolo2019universal} }}
  \label{rot-curves}
\end{figure}

Our theory can calculate the rotation curves for any prescribed surface density $\Sigma_B(s)$ including exponential disks. 
For example, the involute given by the curve $z = - \frac{Al_0}{e} \exp \left(-\frac{r}{l_0}\right)$, depicted in Fig.~\ref{construction}, has a corresponding  surface density 
\begin{align}
\Sigma_B(s) & = \frac{4\Sigma^*}{V_0} \cdot \begin{cases} \quad (1 + A^{-2} e^2 s^2/l_0^2)^{-3/2}, & s \leq l_0, \\
\quad \left(1+A^{-2} e^{2s/l_0}\right)^{-3/2}, & s \geq l_0. \end{cases}
\label{mod-exp}
\end{align}
This family of densities interpolates between a Kuzmin disk ($A l_0 = e z_0$, $A \to 0$) and an exponential disk with a small core ($l_0 \log A \simeq z_0$, $l_0 \to 0$).

Fig.~\ref{rot-curves} shows the numerically obtained rotation curves,  with $A = 1, \Pi_0 = \max(g_{\text{bar}})/a_0 \approx 0.416$ and $A = 9, \Pi_0 \approx 0.724$.  The axes are non-dimensionalized by $(r_0,v_{\text{Kep}})$, 
where $v_{\text{Kep}} = \max(v_{\text{disk}})$ is the peak Keplerian velocity for the density $\Sigma_B$ in~\eqref{mod-exp}, and $r_0$ is set so the peak occurs at $r=2.15\, r_0$ (the peak location for exponential disks). These numerically obtained curves are in excellent agreement with observations \cite{DiPaolo2019universal}.

\begin{figure} 
  \centering
  \includegraphics[width=0.45 \textwidth]{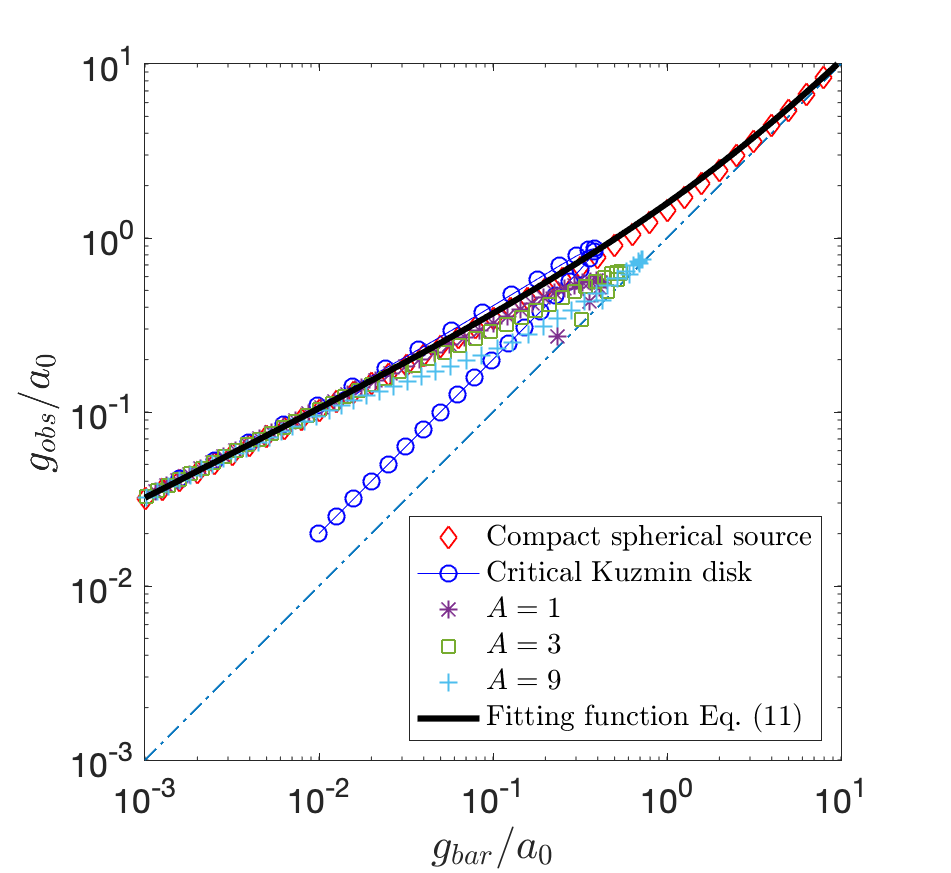} 
  \caption{The radial acceleration relation (RAR) in our model for various distributions of matter.   $g_{\text{obs}}$ and $g_{\text{bar}}$ are  the total gravitational acceleration and the baryonic contribution respectively. No parameters are fit: $g_{\dag} = a_0 = 2 \pi G \Sigma^*$.  Compare Fig.~1 in Ref.~\protect{\cite{DiPaolo2019RAR}}. }
  \label{fig:rar}
\end{figure}

In Fig.~\ref{fig:rar}, we plot the RAR for our theory applied to various matter distributions ($V_0$ is set to 4) and compare with the fit
\begin{equation}
 g_{\text{obs}} = \frac{g_{\text{bar}}}{1- e^{-\sqrt{g_{\text{bar}}/g_{\dag}}}} 
 \label{eq:fitting}
\end{equation} 
for the choice $g_{\dag} = a_0$ \cite{Lelli2017onelaw}. 
Note that, for rotation supported systems, the resulting RAR has two branches, most clearly seen for the critical Kuzmin disk. In the absence of a bulge/central mass,  the baryonic contribution to the acceleration $g_{\text{bar}}$ has a peak value $g_{\max}$ at a few scale-lengths. For $g < g_{\max}$ there are two values of $r$, one on either side of the peak, with $g_{\text{bar}}(r) = g$ and (generically) {\em different} values of $g_{\text{obs}}$. Our theory therefore gives two branches for the RAR in agreement with recent observations for dwarf disk and LSB galaxies~\cite{DiPaolo2019RAR}. Our theory therefore is incompatible with {\em any} MOND rule $g_{\text{obs}} =  \mu(\frac{g_{\text{bar}}}{a_0}) g_{\text{bar}}$ \cite{Milgrom_MOND_1983}. Indeed, the function in~\eqref{eq:fitting} only fits one branch of the RAR. The other branch, $g_{\text{obs}} \simeq g_{\text{bar}} \ll a_0$, gives a possible explanation for galaxies containing very little dark matter \cite{vanDokkum2018galaxy,Famaey2018DF2}. 

%
%

\section{Discussion}

We have proposed a novel theory, derived from an action principle, by combining ideas from pattern formation with a few principles derived from observations, namely the BTFR and the constancy of the central surface density of DM halos.  We have introduced an additional ``dark field" $\psi$ that plays the role of DM. In our theory, no structures are formed on scales smaller than $k_0^{-1}$. The resulting 
``DM halo" is therefore naturally cored in contrast to the cuspy halos formed by CDM. 

Our theory is based on universal equations for pattern formation and can thus describe a variety of instability generating mechanisms. Our favored interpretation is that $\psi$ is the order parameter for a broken translational symmetry, with a characteristic scale $k_0^{-1}$ determined by the distribution of baryons. One realization of this idea, that we are investigating further, is to set $k_0$ as the most unstable wavenumber in  the stability analysis that leads to Toomre's criterion \cite{galactic-dynamics}, thereby directly relating the `baryonic instability' of a rotating disk to the pattern instability that produces the `dark halo' in our framework. 

 Ours is an effective, long wave theory, applicable on scales $\gtrsim k_0^{-1}$, rather than a fundamental theory, since the parameters $\rho_0$ and $k_0$ in~\eqref{Lagrangian} are not universal, but  explicitly depend on the baryonic mass $M_B$ of the host galaxy. This nonlocality is to be expected 
 \cite{Deffayet2011nonlocal}. Our theory is ``minimally" nonlocal through the dependence of its action on  a single global quantity $M_B$. 
 
With no additional fitting parameters, our parsimonious theory predicts a wealth of  observed regularities and scaling laws for isolated, quasi-steady systems, including the RAR, Renzo's rule and the existence of the Freeman limit.  Our model recovers the observed diversity in the rotation curves of LSB disk galaxies, and the inferred ``DM halos", in their fine details. 

Let us reemphasize the two distinct sources for the new effects arising in our theory. First, the curvature of the phase surfaces contributes an additional energy, consistent with ``cored" halos. The resulting gravitational acceleration dominates the baryonic contribution $g_{\text{bar}}$ at large distances and flattens the rotation curves. Second, for disk galaxies, the phase surfaces are spheroidal rather than spherical and thus generate a phase grain boundary on the galactic plane. This additional source yields the third relation in Eq.~\eqref{galaxy}, linking $\theta(s)$, the angle between the phase surfaces and the galactic plane, to the density of the disk, $\Sigma(s)$,  thus providing a natural explanation for the disk-halo connection in galaxies \cite{Sancisi2004visible,DiPaolo2019universal}.  

In ongoing work we are: 
(1) Using a perfect fluid in place of the dust Lagrangian in Eq.~\eqref{Lagrangian} 
to model pressure supported systems. (2) Addressing the Ostrogradski instability for the action in Eq.~\eqref{Lagrangian}, by breaking Lorentz invariance as in Ho\v{r}ava gravity \cite{Horava2009quantum,Blas_Consistent_2010}, to get a Lagrangian only involving first order time derivatives in a preferred slicing. (3) Explicating the role of primordial/dynamical defects in $\psi$ for baryon acoustic oscillations/structure formation. (4) Varying $k_0$ in space for studying clusters of galaxies and compare the effective dark matter in our model with what is inferred from gravitational lensing. (5) Evaluating the consequences  of the potential instability of the PGB when the angle $\theta(s)$ becomes too sharp  \cite{newell1996defects}. 

Galaxy formation is a complex process, and involves a great many effects \cite{Dalcanton1997Formation,Wechsler2018Connection} not included in our simple model. In cosmological contexts galaxies are ``nonlinear", with the implication sometimes being that theorists can build models that describe the very largest scales of the universe, and not be too concerned with tensions between theories and observations, or ``unexplained" regularities/scaling laws, on small ``nonlinear" scales \cite{Famaey2012MOND}. We disagree with this point of view.

  We contend that the robust scaling relations satisfied by galaxies are not ``accidental" and require robust explanations. Our model offers conceptual insight into these relations by demonstrating how a generic mechanism for coupling the dark field $\psi$, through its defects, to the baryonic density $\rho_B$ leads to self-organization. Additionally, our model is a useful technical tool. As we discuss elsewhere, we can embed this theory into a Renormalization group (RG) through scaling transformations for the quantities in~\eqref{galaxy} (See also \cite{Milgrom2009MOND}). Under the RG flow, the evolute $\gamma$ degenerates to a single point, and the one parameter family of critical Kuzmin disks are fixed points. The critical Kuzmin disks therefore ``shepherd" the behavior of rotation supported disk galaxies. In particular, scaling relations for the Kuzmin disks, like the RAR we obtain in~\eqref{kuzmin-pot}, {\em ought to}  hold approximately for general disk galaxies, as in Fig.~\ref{fig:rar}. Also, the phase contours for exponential disks are (approximately) spherical caps $\psi \approx (r^2 + (z+z_0)^2)^{1/2}$, as illustrated in Fig.~\ref{construction}, implying that the Kuzmin disk solutions approximate the ``dark halos" of disk galaxies \cite{Brada1995Exact}. This {\em universality} \cite{Kadanoff1990scaling} justifies our use of simplified physical models, with relatively few ingredients, in our initial attempt to understand self-organization in galaxies. 
 
Our results underscore the need for going beyond CDM and incorporating baryonic feedback or other physical processes, that couple baryonic and dark matter, in order to explain  observed phenomenology on galactic scales \cite{Navarro2017MDARfull,Dutton2019origin}. To the extent our `universal' model reproduces observations, it constrains theories with  baryonic feedback since  Eq.~\eqref{Lagrangian} should emerge as a limit theory in the appropriate scaling regime. 

Beyond discriminating among CDM + feedback theories, our model goes further in suggesting that  `dark matter' can arise as a collective, emergent phenomenon ({\em cf.} \cite{verlinde2017}) and not only as an yet undiscovered particle. The viability of this idea merits further study from an astrophysical viewpoint {\em and} from the viewpoint of complex systems/pattern formation.

\section*{Acknowledgments}
We are grateful to Peter Behroozi and Stacy McGaugh for many illuminating discussions. SCV was partially supported by the Simons Foundation through award 524875 and by the National Science Foundation through award DMR-1923922.  

%





\end{document}